\documentclass{webofc}

\usepackage[varg]{txfonts}   
\usepackage{hyperref}
\usepackage{url}
\hypersetup{colorlinks=true,citecolor=blue,urlcolor=blue,linkcolor=blue}

\begin{document}

\title{Probing momentum dependence of hydrodynamization in heavy-ion collisions}

\author{\firstname{Akihiko} \lastname{Monnai}\inst{1}\fnsep\thanks{\email{akihiko.monnai@oit.ac.jp}}
}

\institute{ 
Department of General Education, Faculty of Engineering, Osaka Institute of Technology, Osaka 535-8585, Japan
}

\abstract{
The fluidity of the hot and dense QCD matter is a key characteristic of the medium created in high-energy heavy-ion collisions. We extend the framework of the relativistic hydrodynamic model to incorporate non-thermal momentum distributions that may emerge during the dynamical evolution of the collision system. Numerical simulations are performed to elucidate the phenomenological implications of these modifications on charged hadrons and direct photons measured in collider experiments. 
}
\maketitle
\section{Introduction}
\label{sec:1}
\begin{figure}[h]
\centering
\includegraphics[width=12.5cm,clip]{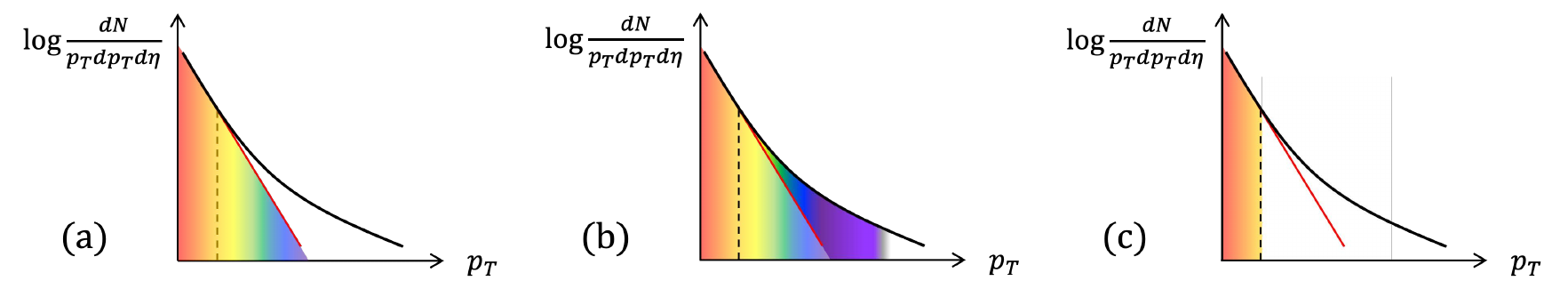}
\caption{Schematic of transverse momentum spectra in nuclear collisions and momentum regions covered by the hydrodynamic model with (a) conventional thermal distribution, (b) Tsallis distribution, and (c) thermal distribution with a momentum cutoff.}
\label{fig:1}
\end{figure}

The relativistic hydrodynamic model has been indispensable for analyzing the quark-gluon plasma (QGP) created in high-energy nuclear collisions at the BNL Relativistic Heavy Ion Collider and CERN Large Hadron Collider. Experimental observations indicate that low-momentum components below approximately $p_T\sim2$-$4$ GeV are thermal and hydrodynamic, while high-momentum tails are non-thermal and perturbative. One often employs a hybrid description of the hydrodynamic model and perturbative QCD to understand the transverse momentum ($p_T$) spectra. 
In this study, we investigate two scenarios of hydrodynamization without conventional thermalization: (i) the `violet' hydrodynamic model \cite{Kyan:2022eqp} where an extended momentum range is treated as hydrodynamic based on Tsallis statistics \cite{Tsallis:1987eu,Tsallis:1999nq}; and (ii) the `red' hydrodynamic model \cite{Monnai:2023evo} where medium-high momentum contributions are excluded from the bulk medium (Fig.~\ref{fig:1}). Using numerical simulations, we elucidate the momentum dependence of hydrodynamization and its effect on flow observables in heavy-ion collisions.

\section{The model}
\label{sec:2}

\subsection{Hydrodynamic model with high-momentum components}
\label{sec:2-1}

We introduce Tsallis statistics to incorporate part of the high $p_T$ tail observed in particle spectra into the hydrodynamic description \cite{Osada:2008sw,Takacs:2019ikb,Kyan:2022eqp}. The conserved charges are neglected for simplicity. Tsallis statistics is a generalization of thermal statistics characterized by the parameter $q$. Fermi-Dirac and Bose-Einstein distributions in $q$-equilibrium are
\begin{eqnarray}
f(E,T,q) = \frac{\exp_q (-E/T)}{1\pm \exp_q (-E/T)}, 
\end{eqnarray}
where $\exp_q (x) = [1+(1-q)x]^{\frac{1}{1-q}}$. They approach thermal distributions in the limit of $q\to 1$. Tsallis distributions exhibit a power-law-like tail structure when $q>1$. 

The energy-momentum conservation $\partial_\mu T^{\mu \nu} = 0$ remains the same. On the other hand, the equation of state is affected by $q$. The pressure is expressed in kinetic theory as  
\begin{eqnarray}
P = \frac{1}{3} \sum_i \int \frac{g_i d^3p}{(2\pi)^3} \frac{\mathbf{p}^2}{E_i} f_i^q , 
\end{eqnarray}
where $i$ is the index for particle species. We consider a hadron resonance gas for the hadronic phase and a parton gas for the QGP phase. They are smoothly connected at the connection temperature $T_c$. Below $T_c$, the pressure is $P(T) = P_\mathrm{had}(T) $. Above $T_c$, it is parametrized as
\begin{eqnarray}
P(T) = P_\mathrm{had}(T_c) + [P_\mathrm{QGP}(T)-P_\mathrm{had}(T_c)] \{1-\exp[-c(T-T_c)]\} , 
\end{eqnarray}
where the constant $c$ is chosen so that the pressure is continuously differentiable.

\begin{figure}[tb]
\centering
\includegraphics[width=5.0cm,clip]{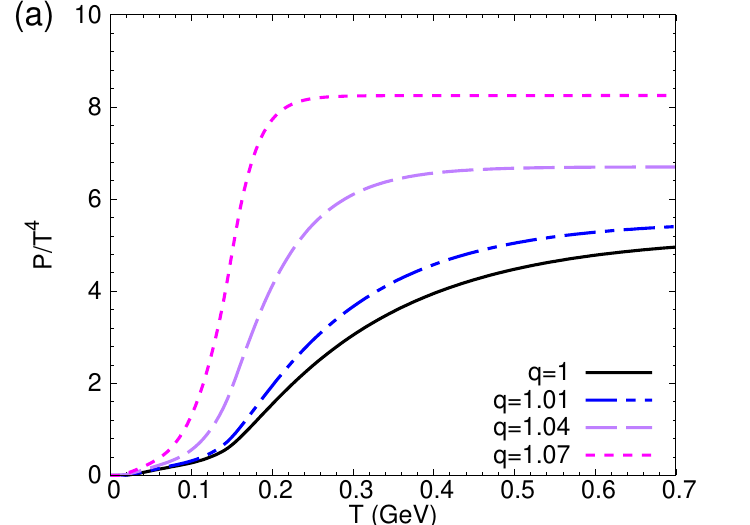}
\includegraphics[width=5.0cm,clip]{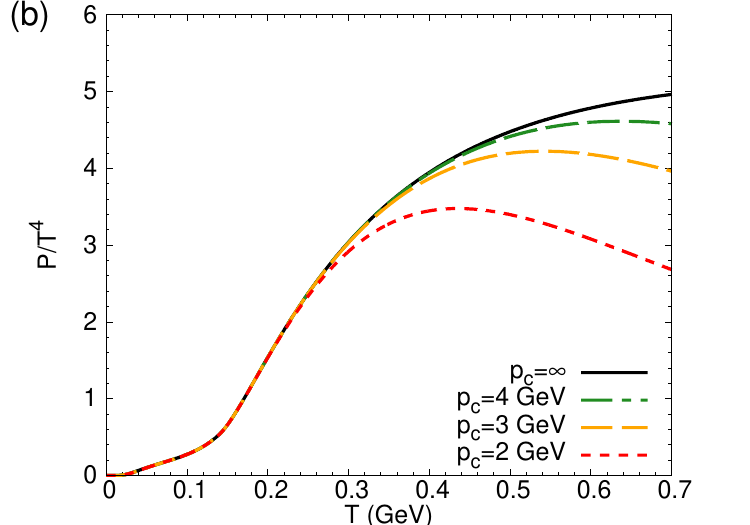}
\caption{The equation-of-state models (a) with Tsallis statistics and (b) with momentum cutoffs.}
\label{fig:2}
\end{figure}

Figure~\ref{fig:2} (a) shows $P/T^4$ for different $q$ at $T_c = 0.14$ GeV when the hadron resonances composed of $u,d,s$ with the mass below 2 GeV and the $N_f=3$ parton gas are considered. It increases with $q$ owing to contributions from higher momentum components. 

The particlization is performed using the Tsallis-extended Cooper–Frye prescription to ensure energy-momentum conservation. 

\subsection{Hydrodynamic model with low-momentum components}
\label{sec:2-2}

The conventional hydrodynamic model assumes that part of the medium-high momentum components are thermalized. They can be excluded from the system by imposing a momentum cutoff $p_c$ on thermal distributions. The pressure is expressed as 
\begin{eqnarray}
P = \pm T \sum_i \int_0^{p_c} \frac{g_i d^3p}{(2\pi)^3} \ln{\bigg[1\pm \exp \bigg(-\frac{E_i}{T} \bigg) \bigg]} , 
\end{eqnarray}
in kinetic theory. We employ the same connection method as in Sec.~\ref{sec:2-1} to bridge the hadron resonance gas and the parton gas pressures. 

The dimensionless pressure for different $p_c$ is shown in Fig.~\ref{fig:2} (b). Lower momentum cutoffs lead to smaller pressure at higher temperatures. On the other hand, they have negligible effects in the hadronic phase when $p_c \geq 2$ GeV. 
This implies that the hydrodynamic evolution is affected mostly at early times, and the effect of the cutoff on the bulk medium at the particlization is, unlike the Tsallis case, relatively small. 

We consider direct photons as a candidate probe sensitive to $p_c$. Prompt photons are parametrized as in Ref.~\cite{Turbide:2003si}. The thermal photon emission rate in the QGP phase with a cutoff is, based on the small-angle approximation \cite{Berges:2017eom, Blaizot:2014jna},
\begin{eqnarray}
E \frac{dR^\gamma_\mathrm{QGP}}{d^3p} = \sum_f \frac{4 e_f^2 }{\pi^2} \alpha_\mathrm{EM} \alpha_s \log \bigg(1+\frac{2.919}{g^2}\bigg)  h(p) f_q(p) \int_0^{p_c} \frac{d^3p'}{(2\pi)^3} \frac{1}{p'} [f_g(p') + f_q(p')], 
\end{eqnarray}
where
$h(p) = \{ 1 - \tanh [ (p-p_c)/\Delta p_c]\}/2$. The rate in the hadronic phase is estimated based on Ref.~\cite{Turbide:2003si} truncated with $h(p)$. The rates are connected around $T_\mathrm{ph}=0.17$ GeV as
\begin{eqnarray}
E \frac{dR^\gamma_\mathrm{th}}{d^3p} = \frac{1}{2} \bigg[ 1 - \tanh\bigg( \frac{T-T_\textrm{ph}}{\Delta T_\textrm{ph}} \bigg) \bigg] E \frac{dR^\gamma_\mathrm{had}}{d^3p} + \frac{1}{2} \bigg[ 1 + \tanh\bigg( \frac{T-T_\textrm{ph}}{\Delta T_\textrm{ph}} \bigg) \bigg] E \frac{dR^\gamma_\mathrm{QGP}}{d^3p}. 
\end{eqnarray}

High-momentum components excluded from the hydrodynamic medium should also emit photons. Here, the emission rate is conjectured by the difference between the thermal rate with and without cutoff. High $p_T$ photons are assumed to have zero anisotropy. 

\section{Numerical results}
\label{sec:3}

We demonstrate the effects of momentum-dependent hydrodynamization using a (2+1)-dimensional inviscid numerical hydrodynamic model. The Monte-Carlo Glauber model is used to construct event-averaged initial conditions for 2.76 TeV Pb+Pb collisions.

\begin{figure}[tb]
\centering
\includegraphics[width=4.2cm,clip]{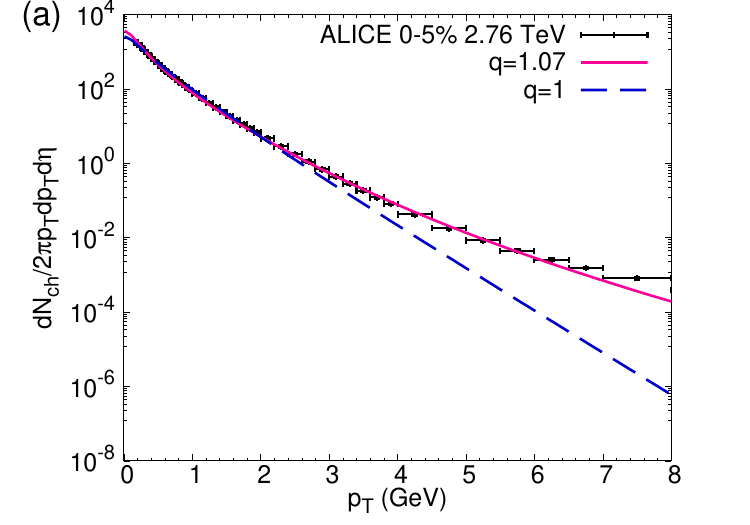}
\includegraphics[width=4.2cm,clip]{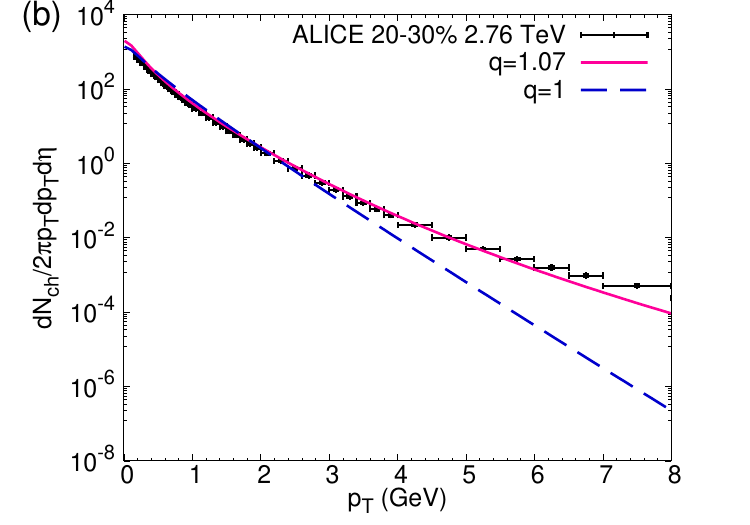}
\includegraphics[width=4.2cm,clip]{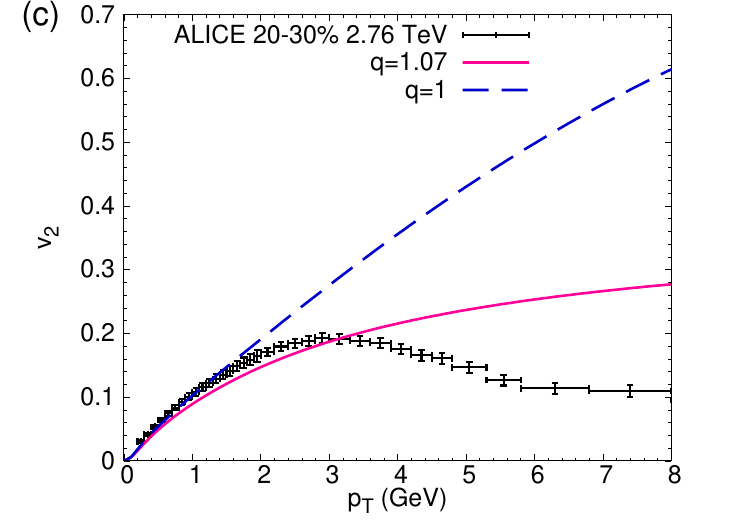}
\caption{Charged particle $p_T$ spectra for (a) 0-5\% and (b) 20-30\% centrality classes and (c) $v_2$ for the 20-30\% centrality class with thermal ($q=1$) and Tsallis ($q=1.07$) distributions.}
\label{fig:3}
\end{figure}

$p_T$ spectra and elliptic flow $v_2$ of charged hadrons from the Tsallis hydrodynamic model are shown in comparison to the ALICE data \cite{ALICE:2012aqc} in Fig.~\ref{fig:3}. The thermal results ($q=1$) describe the $p_T$ spectra up to around 2-3 GeV, while the Tsallis results at $q=1.07$ are compatible with the data up to 6-8 GeV for 0-5\% and 20-30\% centrality classes. On the other hand, the elliptic flow $v_2$ suggests a narrower applicability range, though event-by-event simulations are required for further quantitative analysis.

\begin{figure}[tb]
\centering
\includegraphics[width=4.2cm,clip]{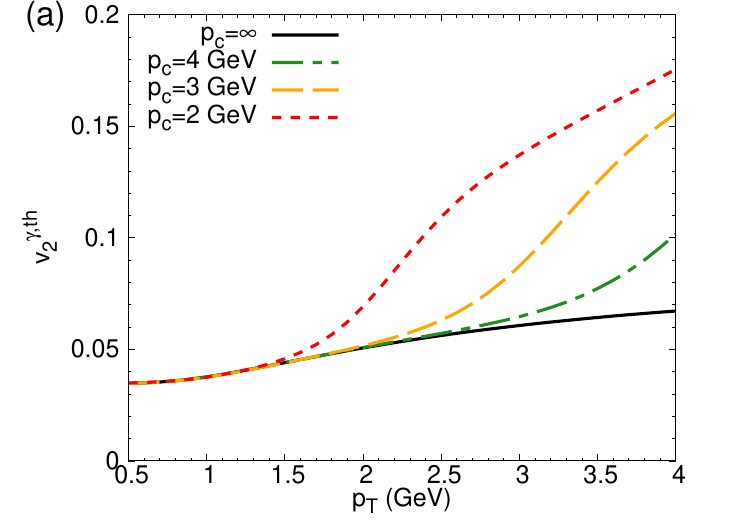}
\includegraphics[width=4.2cm,clip]{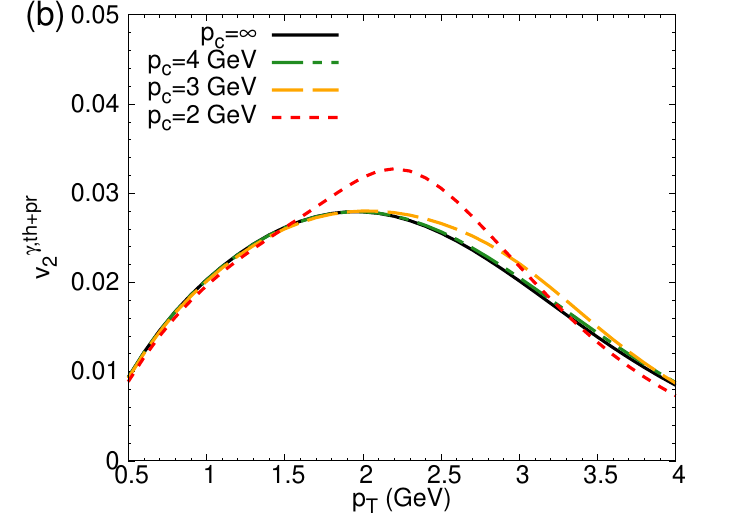}
\includegraphics[width=4.2cm,clip]{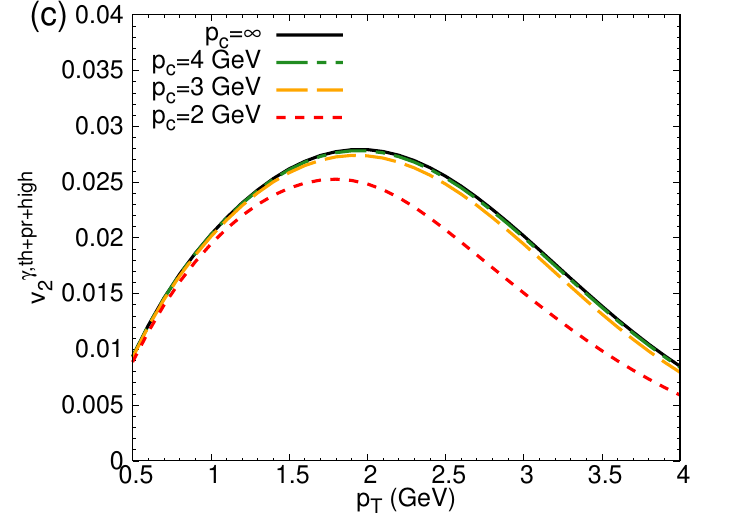}
\caption{Elliptic flow of (a) thermal photons, (b) thermal and prompt photons, and (c) direct photons with different momentum cutoffs.}
\label{fig:4}
\end{figure}

Figure~\ref{fig:4} shows the direct photon elliptic flow $v_2$ corresponding to the 0-20\% centrality class. Here, direct photons are defined as the sum of thermal, prompt, and high-$p_T$ photons. Thermal photon $v_2$ is enhanced by the momentum cutoff since the contributions from the early stages of nuclear collisions -- where the momentum anisotropy is still underdeveloped -- are suppressed. Adding prompt photons suppresses $v_2$ as they have no anisotropy but the presence of a momentum cutoff still enhances $v_2$. If one introduces high $p_T$ photons as described in Sec.~\ref{sec:2}, the elliptic flow is further suppressed and becomes smaller than that without a cutoff. It should be noted that whether the momentum cutoff enhances or suppresses direct photon $v_2$ depends on the details of high $p_T$ photon production mechanisms.

\section{Summary and conclusion}
\label{sec:4}

We have developed extended frameworks of the relativistic hydrodynamic model based on non-thermal phase-space distributions to investigate momentum-dependent hydrodynamization. First, we have introduced Tsallis statistics to incorporate part of the high $p_T$ tail into the hydrodynamic model. This leads to larger pressure for a given temperature. Numerical simulations indicate that the momentum range of applicability can be extended when $q=1.07$. Second, we have truncated the distribution in momentum space to exclude medium-high momentum components. Using a modified equation of state, direct photons are estimated. The results suggest that a momentum cutoff can leave non-trivial imprints on direct photon elliptic flow. Future prospects include the introduction of viscosity and event-by-event analyses.

\section*{Acknowledgments}
\label{sec:a}
Part of this work is based on collaboration with K. Kyan. The author thanks M. Kitazawa, K. Murase, A. Ohnishi, H. Suganuma, and S. Yoshikawa for valuable discussions. The work of A.M. was supported by JSPS KAKENHI Grant Numbers JP19K14722 and JP24K07030.

\end{document}